\documentclass[10pt,twocolumn,english,showpacs]{revtex4}
\usepackage{times}
\usepackage[T1]{fontenc}
\usepackage[latin1]{inputenc}
\usepackage{graphicx}
\usepackage{amssymb}

\makeatletter

\usepackage{babel}
\makeatother
\begin{document}

\title{Bose-Einstein condensation in an optical lattice}
\author{P. B. Blakie$^1$ and Wen-Xin Wang$^{1,2}$}

 \affiliation{1. Jack Dodd Centre for Photonics and Ultra Cold Atoms, Department of
Physics, University of Otago, Dunedin, New Zealand\\ 2. Physics Department, Petroleum University of China (East), Dongying,
Shandong, China 257062}


\email{bblakie@physics.otago.ac.nz}

\begin{abstract}
In this paper we develop an analytic expression for the critical temperature
for a gas of ideal bosons in a combined harmonic lattice potential,
relevant to current experiments using optical lattices. We give corrections
to the critical temperature arising from effective mass modifications
of the low energy spectrum, finite size effects and excited band states.
We compute the critical temperature using numerical methods and compare
to our analytic result. We study condensation in an optical lattice
over a wide parameter regime and demonstrate that the critical temperature
can be increased or reduced relative to the purely harmonic case by
adjusting the harmonic trap frequency. We show that a simple numerical
procedure based on a piecewise analytic density of states provides
an accurate prediction for the critical temperature. 
\end{abstract}
\maketitle

\section{Introduction}

Bosonic atoms confined in optical lattices have proven to be a versatile
system for exploring a range of physics \cite{Anderson1998a,Burger2001a,Greiner2002b,Hensinger2001a,Morsch2003a,Orzel2001a,Spielman2006a}, exemplified by the superfluid to Mott-insulator transition \cite{Greiner2002a,Jaksch1998a}.
In the superfluid limit a condensate exists in the system and  experiments
have explored its properties, such coherence \cite{Orzel2001a,Greiner2001a,Morsch2002a,Spielman2006a},
collective modes \cite{Fort2003}, and transport \cite{Fertig2005,Burger2001a,Fallani20046}.
While several experiments have considered the interplay of the condensate
and thermal cloud at finite temperatures \cite{Fort2003,Greiner2001a},
the nature of the condensation transition itself remains to be examined.

For the 3D Bose gas significant theoretical attention has been given
to the condensation transition. The ideal uniform gas has the well-known
critical temperature $T_{c0}\sim(N/V)^{2/3}$, although it is only
recently that consensus has been reached on how s-wave interactions
shift this result \cite{Baym1999a,Baym2001a,Arnold2001c,Kashurnikov2001a,Davis2003a,Andersen2004a}.
In the experimentally relevant harmonically trapped case the ideal
transition temperature scales as $T_{c0}\sim\bar{\omega}N^{1/3}$
in the thermodynamic limit. Finite size \cite{Dalfovo1999a} and
interaction effects (at the meanfield level) \cite{Giorgini1996a}
give important corrections, and including them is necessary to obtain
good agreement with experiment \cite{Gerbier2004a} (also see \cite{Zobay2004b,Davis2006a}).

While the occurrence of condensation in a lattice is hardly surprising
(when interactions are small), there are few theoretical
predictions for the condensation temperature or behaviour.
 For the idealized case of a (uniform) translationally invariant lattice Kleinert
\emph{et al.} \cite{Kleinert2004} have made predictions that a re-entrant
phase transition will be observed with varying interaction strength. Ramakumar \emph{et al.} \cite{Ramakumar2005b} have also examined interaction effects in the translationally invariant lattice, and have explored the critical temperature dependence on lattice geometry.

\begin{figure}
\includegraphics[width=2in]{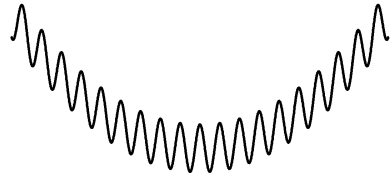}
\caption{\label{fig:CombLatt} Schematic diagram showing the 
combined harmonic lattice potential considered in this
paper.}
\end{figure}

 In experimentally produced optical lattices the periodic potential is always accompanied by a harmonic potential, produced by the focused light fields used to make the lattice and sometimes enhanced by magnetic trapping (e.g. see \cite{Gerbier2007a}). 
We refer to this experimentally realistic potential as the \emph{combined harmonic lattice potential} (see Fig. \ref{fig:CombLatt}). We are aware of two numerical studies that have considered finite temperature condensation in the combined potential \cite{Wild2006,Ramakumar2007a}. Wild \emph{et al.} \cite{Wild2006} have considered a quasi-1D system and examined the effect of interactions
on the transition temperature using a meanfield approach. Ramakumar
\emph{et al.} \cite{Ramakumar2007a} used numerical studies to examine
condensation and thermal properties for the ideal gas limit. All of
these studies \cite{Kleinert2004,Wild2006,Ramakumar2005b,Ramakumar2007a} have used a tight-binding description (or Bose-Hubbard
model) that neglects the role of higher vibrational bands, and can
only be applied when the lattice is sufficiently deep and the atoms
are sufficiently cold.  Going beyond the tight-binding approximation  Zobay \emph{et al.} \cite{Zobay2006b} have used meanfield and renormalization treatments to consider the effects of interactions in a uniform system with a weak one-dimensional translationally invariant lattice  (depth less than a recoil energy).

We also note several studies showing how adiabatic variations of the lattice depth
might be used to prepare a condensate or reversibly condense the system
 \cite{Olshanii2002a,Blakie2004b} and recent papers debating
the use of interference peaks  as a signature of condensation \cite{Diener2007a,Gerbier2007a,Yi2007a}.

The central difficulty in calculating the properties of a quantum
gas in the combined potential is that the spectrum has a rich and
complex structure. Several articles have considered aspects of this
system \cite{Hooley2004a,viverit2004a,Rigol2004b,Rey2005a} for the
case of one or two spatial dimensions. The first study we are aware
of is a tight-binding description of ultra-cold bosons by Polkovnikov
\emph{et al}. \cite{Polkovnikov2002a}. Refs. \cite{Hooley2004a,Rigol2004b}
have made detailed studies of the combined potential spectrum (also
within a tight-binding description), and more recently closed-form
solutions were given by Rey \emph{et al}. \cite{Rey2005a}. In Refs.
\cite{viverit2004a,Ruuska2004a} an ideal gas of fermions in a 1D
combined potential was examined without making the tight-binding approximation.
All of these studies have confirmed that, for appropriate parameter
regimes, parts of the single particle spectrum will contain localized
states. This is in contrast to the translationally invariant system,
where inter-atomic interactions or disorder are needed for localization
to occur (e.g. see \cite{Jaksch1998a,Buonsante2007a}). Experiments
with ultra-cold (though non-condensed) bosons \cite{Ott2004b} have
provided evidence for these localized states.

In this paper we present a theory for an ideal Bose gas in a three
dimensional combined harmonic lattice potential. Our treatment includes
excited bands, and is thus valid at high temperatures where tight-binding
descriptions fail. Central to our approach is the division of the
spectrum into two regions: (1) A low energy region consisting of extended
oscillator-like states, modified from those of the harmonic potential
by the low energy effective mass. (2) A high energy region containing  localized states
that have an energy related to the local potential energy, and also
includes modes in the first vibrational excited bands. The division
between these regions is made by the use of an effective Debye energy
for the system.

In current experiments with optical lattices inter-particle interactions
are typically important, at least in determining the near zero temperature
manybody ground state. Interaction effects near the critical region have yet
to be examined for the combined potential (although the quasi-1D case
is examined in \cite{Wild2006}), and our ideal gas results will
be a useful basis for comparison with future studies. 

We begin in Section \ref{sec:Formalism} by introducing relevant energy
scales and describe an approximate analytic spectrum and density of
states in the combined potential. In Sec. \ref{sec:Results} we compare
our analytic density of states to the results of full numerical calculations
to justify the validity regime of our analytic approach. We then derive
an analytic approximation to the critical temperature in the combined
potential and calculate corrections resulting from the low energy
spectrum and the effects of excited bands. Those results are compared
to full numerical calculations to assess their accuracy and validity.
Finally in Sec. \ref{sec:GenRes} we present some general numerical
results for the condensation phase diagram in the combined potential.
We show how varying the harmonic confinement can be used to raise
or lower the transition temperature relative to the pure harmonically
trapped case.

\section{Formalism\label{sec:Formalism}}

\subsection{Single particle Hamiltonian}

We consider the case of a single particle Hamiltonian of the form

\begin{equation}
H=\frac{p^{2}}{2m}+\sum_{j=1}^{3}\left[V_{j}\sin^{2}\left(\frac{bx_{j}}{2}\right)+\frac{1}{2}m\omega_{j}^{2}x_{j}^{2}\right],\label{eq:Hsp}\end{equation}
where $b$ is the reciprocal lattice vector, and $\{ V_{1},V_{2},V_{3}\}$
are the lattice depths in each direction. It is conventional to define the recoil energy
 $E_{R}=\hbar\omega_{R}=h^{2}/8ma^{2}$ as an energy scale for
specifying the lattice depth, where $a=2\pi/b$ is the direct lattice
vector. Properties of the single particle spectrum have been discussed
by several authors, e.g. see Refs. \cite{Blakie2007c,Rey2005a,Rigol2004b,Hooley2004a,viverit2004a}.
Here we use the results of these studies to suggest an approximate
(piecewise) analytic density of states for the combined lattice appropriate
for determining the critical temperature. 

We begin in the next subsection  by defining a set of useful quantities
that will be crucial for developing approximations to the spectrum
in different regimes. These quantities can be determined from solutions
of the much simpler translationally invariant lattice (i.e. Eq. (\ref{eq:Hsp})
with all $\omega_{j}=0$) or from analytic approximations valid in
the tight-binding regime.

\subsection{Energy scales from the translationally invariant lattice\label{sub:EnergyScales}}
\begin{figure}
\includegraphics[width=3.4in]{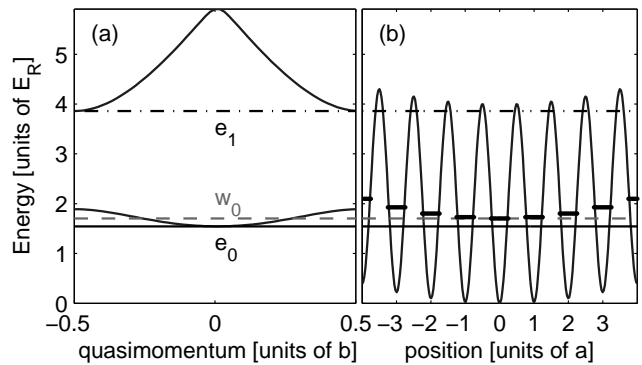}
\caption{\label{Fig:Escales}Schematic diagram of important energy scales.
(a) Band structure of a translationally invariant 1D lattice of depth
$4E_{R}$. Energy scales $e_{0}$ (black horizontal line) , $w_{0}$
(grey dashed line) and $e_{1}$ (dash-dot line) identified (see text).
(b) Correspondence of these energy scales to the combined potential.
Small thick horizontal lines indicate the energies of ground band
localized states.}
\end{figure}
 A one-dimensional depiction of the important energy scales is given in
Fig. \ref{Fig:Escales}. There we show the one dimensional band structure
[Fig. \ref{Fig:Escales}(a)], and indicate several energies that
we discuss further below.

\subsubsection{Bloch state parameters}
The quantity $e_{n}$ refers to the minimum (Bloch state) energy of
the band with $n$-vibrational quanta, and in the full 3-dimensional
case we will use the notation $e_{nj}$ to denote the particular $n$-th
excited band by specifying additional quantum number(s) $j$. Here
we will only refer to a few of these band minimum energies: $e_{0}$
is the ground state energy of the translationally invariant lattice
and gives a lower bound for the ground state energy when the harmonic
trap is added; $e_{1j}$ is the lowest energy of the first vibrational
excited state, with the quantum number $j=1,2,3$ used indicate that
the vibrational excitation is directed along the $x_{j}$-direction;
We use $e_{2}$ to indicate the energy above which higher excited
bands become accessible %
\footnote{As we only use $e_{2}$ to establish a validity condition for our
theory   we dispense with any additional quantum numbers to label
the second excited band. Generally we take $e_{2}$ to be the lowest
energy at which a second excited band state becomes accessible.%
}.

\subsubsection{Wannier state parameters}
We define $w_{0}$ as the energy of a localized Wannier state in the
ground band. Wannier states are defined as a Fourier transform of
the ground band Bloch states (e.g. see \cite{Jaksch1998a,Blakie2004a}),
and as such its energy is the mean energy of all the ground band Bloch
states (see Fig. \ref{Fig:Escales}(a)). The tunneling between neighbouring
Wannier states in the $x_{j}$-direction is characterized by the tunneling
matrix element $J_{j}$. This is given by the Fourier transform of
the ground band Bloch dispersion relation along direction $x_{j}$.

\subsubsection{Effective mass}
Intermediate between the extended Bloch states and localized Wannier
states we will need to describe finite extent wavepackets in the
ground band. A convenient quantity for doing this is the effective
mass at zero quasimomentum, $m_{j}^{*}$, defined as \begin{equation}
\frac{1}{m_{j}^{*}}=\frac{1}{\hbar^{2}}\left(\frac{\partial^{2}e_{0}(\mathbf{q})}{\partial q_{j}^{2}}\right)_{\mathbf{q=0}},\label{eq:effmass}\end{equation}
where $e_{0}(\mathbf{q})$ is the dispersion relation of the ground
band and $\mathbf{q}$ is the quasimomentum. We note that the effective
mass may be different along each direction.

\subsubsection{Tight-binding expressions}
All of the above quantities are easily obtained from calculations
of the translationally invariant lattice or equivalently from the
well-known properties of the Matthieu functions. However the tight-binding
limit, which should be applicable when $V_{j}\gtrsim5E_{R}$, yields
several simple analytic expressions for these quantities. In the appendix
of Ref. \cite{Blakie2007c} an approximation for the energies of
the excited bands is developed using a harmonic oscillator approximation.
Using those results we obtain $e_{0}\approx\left(\sum_{j}\sqrt{V_{j}E_{R}}\right)-\frac{3}{4}E_{R},$
$e_{1j}\approx(2\sqrt{V_{j}E_{R}}-E_{R})+e_{0}$, and $e_{2}\approx4\sqrt{\min\{ V_{j}\}E_{R}}-3E_{R}+e_{0}.$
The tunneling matrix element can also be calculated using the harmonic
oscillator approximation, giving $J_{j}\approx(4/\sqrt{\pi})(V_{j}/E_{R})^{3/4}\exp(-2\sqrt{V_{j}/E_{R}})E_{R}$.
In the tight-binding limit the ground band dispersion relation is
approximately given by $e_{0}(\mathbf{q})=\sum_{j}4J_{j}\sin^{2}(q_{j}a/2)+e_{0},$
where $\mathbf{q}$ is the quasimomentum. From this we obtain expressions
for the Wannier energy $w_{0}=e_{0}+\sum_{j}2J_{j}$, and the effective
mass $m_{j}^{*}=\hbar^{2}/2J_{j}a^{2}$.

\subsection{Spectrum and density of states in the combined harmonic lattice potential }

\subsubsection{Low energy spectrum ($\epsilon<E_{{\rm LE}}$)\label{sub:LowE}}

The low energy states in the lattice are extended wavepackets, with
a harmonic oscillator envelope. Indeed, the spectrum is that of a
harmonic oscillator but with the frequency modified by the effective
mass, $\omega_{j}^{*}=\sqrt{m/m_{j}^{*}}\omega_{j}$ \cite{Rey2005a},
i.e.\begin{equation}
\epsilon_{{\rm LE}}(\mathbf{n})=e_{0}+\sum_{j=1}^{3}\hbar\omega_{j}^{*}\left(n_{j}+\frac{1}{2}\right),\label{eq:eLE}\end{equation}
where the $\{ n_{j}\}$ are non-negative integers. This low energy
description is valid for quantum numbers in the range $0\le n_{j}\le\mathcal{N}_{j}$
where
\begin{equation}
\mathcal{N}_{j}\equiv4\sqrt{J_{j}/m\omega_{j}^{2}a^{2}},
\end{equation}
(see \cite{Rey2005a}) as for values of $n_{j}$ greater than $\mathcal{N}_{j}$
the states become localized (see below).

The density of states for these modes is given by\begin{equation}
g_{{\rm LE}}(\epsilon)=\frac{(\epsilon-e_{0})^{2}}{2\hbar^{3}\overline{\omega^{*}}^{3}},\quad e_{0}\le\epsilon<E_{{\rm LE}},\label{eq:gLE}\end{equation}
where $\overline{\omega^{*}}=\sqrt[3]{\omega_{1}^{*}\omega_{2}^{*}\omega_{3}^{*}}$
is the geometric mean of the effective trap frequencies. The boundary
of the rectangular region of $\{ n_{j}\}$-space, where the low energy
description is valid, does not correspond to a well-defined energy
cutoff. We introduce an effective Debye energy, $E_{{\rm LE}}$, such
that a total of $\mathcal{N}_{1}\mathcal{N}_{2}\mathcal{N}_{3}$ low
energy states would lie below this energy. A simple calculation yields
\begin{equation}
E_{{\rm LE}}=4\sqrt[3]{6}\hbar\sqrt{\frac{\bar{J}}{\overline{m^{*}}a^{2}}}+e_{0},\label{eq:ELE}\end{equation}
 where $\bar{J}$ and $\overline{m^{*}}$ are the geometric means
of the tunneling matrix elements ($J_{j}$) and effective masses ($m_{j}^{*}$)
respectively. Since $E_{{\rm LE}}$ depends on $\bar{J}$ it is exponentially
suppressed towards $e_{0}$ as the lattice depth increases.

The ground state energy of the combined potential, corresponding to
the state in which the condensate forms, is given by Eq. (\ref{eq:eLE})
with $n_{1}\!=\! n_{2}\!=\! n_{3}\!=\!0,$ i.e.\begin{equation}
\epsilon_{g}=e_{0}+\frac{1}{2}\sum_{j}\hbar\omega_{j}^{*}.\label{eq:Eg}\end{equation}
We see that the effect of the harmonic confinement is to shift the
ground state energy upward from that of the translationally invariant
lattice i.e, $e_{0}$. However, $\epsilon_{g}$ will still be less
than $w_{0}$ if the harmonic potential is less confining than a single
lattice site.

\subsubsection{Localized spectrum ($\epsilon\ge E_{{\rm LE}}$)\label{sub:Loc}}

The next part of the spectrum consists of localized states, arising
because the offset in potential energy between lattice sites near
the classical turning point exceeds the respective tunneling matrix
element. The nature of these states and the derivation of their respective
density of states is treated fully in Ref. \cite{Blakie2007c}, but
we briefly summarize those results here. 

The energies of the localized states are given by the local potential
energy \begin{equation}
\epsilon_{{\rm L}0}(\mathbf{n})=\frac{1}{2}ma^{2}(\omega_{1}^{2}n_{1}^{2}+\omega_{2}^{2}n_{2}^{2}+\omega_{3}^{2}n_{3}^{2})+w_{0},\label{eq:EL0}\end{equation}
where $\{ n_{j}\}$ are (positive and negative) integers that specify
the site where the state is localized. As these states localize to
approximately a single lattice site, their energy offset from the
lattice site minimum (i.e. $\frac{1}{2}ma^{2}\sum_{j}\omega_{j}n_{j}^{2}$)
is given by the Wannier energy $w_{0}$. Schematically these states
are indicated in Fig. \ref{Fig:Escales}(b) as horizontal rungs in
each lattice site (recalling that for $\epsilon_{{\rm L0}}(\mathbf{n})\lesssim E_{{\rm LE}}$
tunneling delocalizes these states).

This description is valid for all energies above $E_{{\rm LE}}$,
however for sufficiently high energy scales additional vibrational
states become available. Here we will also approximate these excited
band states using a localized description, i.e. \begin{equation}
\epsilon_{{\rm L}1j}(\mathbf{n})=\frac{1}{2}ma^{2}(\omega_{1}^{2}n_{1}^{2}+\omega_{2}^{2}n_{2}^{2}+\omega_{3}^{2}n_{3}^{2})+e_{1j},\quad j=1,2,3\label{eq:EL1j}\end{equation}
where we have approximated the zero point energy of these states as
$e_{1j}$. Note that because the vibrational excitation may be directed
along any coordinate direction we have three first excited bands to
include.

The density of states for the spectra given in Eqs. (\ref{eq:EL0})
and (\ref{eq:EL1j}) is \begin{equation}
g_{{\rm Loc}}(\epsilon)=g_{0}(\epsilon-w_{0})+\sum_{j=1}^{3}g_{0}(\epsilon-e_{1j}),\quad E_{{\rm LE}}\le\epsilon<e_{2},\label{eq:gLoc}\end{equation}
where \begin{equation}
g_{0}(\epsilon)=\frac{16}{\pi^{2}}\left(\frac{\omega_{R}}{\hbar\bar{\omega}^{2}}\right)^{3/2}\sqrt{\epsilon}\,\theta(\epsilon),\label{eq:g0}\end{equation}
 $\bar{\omega}=\sqrt[3]{\omega_{1}\omega_{2}\omega_{3}}$, (e.g. see
\cite{Blakie2007c,Kohl2006a}), and $\theta(\epsilon)$ is the unit
step function. We also note that the case of a general (non-separable
lattice) has the same density of states if we instead identify $E_{R}=h^{2}/8mV_{c}^{2/3},$
where $V_{c}=|\mathbf{a}_{1}\cdot(\mathbf{a}_{2}\times\mathbf{a}_{3})|$
is the unit cell volume and $\{\mathbf{a}_{1},\mathbf{a}_{2},\mathbf{a}_{3}\}$
are the direct lattice vectors. 

The localized states description of the first excited band is the
most severe approximation we make for the combined potential spectrum,
particularly because the lowest energy states of the excited bands
will also be harmonic oscillator-like. For deep lattices the tunneling
rates for the ground and excited bands are small and the localized
description improves. For the theory we develop here, the first excited
bands are assumed to be a rather large energy scale compared to the
critical temperature and this approximation should be adequate.

\subsubsection{Bare oscillator states ($\epsilon>E_{{\rm bare}}$)\label{sub:BareOsc}}

At sufficiently high energy scales the lattice has only a small effect
on the energy eigenstates and the spectrum crosses over to bare oscillator
states. This cross-over occurs when the single particle energies exceed
the lattice depth which in 3D we can take as the sum of the lattice
coefficients $E_{{\rm bare}}=\sum_{j}V_{j}$. The bare oscillator
spectrum is of the form given in Eq. (\ref{eq:eLE}) but with the
bare trap frequencies, i.e.\begin{equation}
\epsilon_{{\rm bare}}(\mathbf{n})=\epsilon_{V}+\sum_{j=1}^{3}\hbar\omega_{j}\left(n_{j}+\frac{1}{2}\right),\label{eq:Ebare}\end{equation}
where $n_{j}$ are non-negative integers. The constant  $\epsilon_{V}=\frac{1}{2}\sum_{j=1}^{3}V_{j}$
is the spatial average of the lattice potential and gives the shift
of the high energy spectrum. The density of states is given by \begin{equation}
g_{{\rm bare}}(\epsilon)=\frac{(\epsilon-\epsilon_{V})^{2}}{2\hbar^{3}\bar{\omega}^{3}},\quad\epsilon>E_{{\rm bare}}.\label{eq:gbare}\end{equation}
For the parameter regimes of interest (lattices with depths greater
than a few recoils) $E_{{\rm bare}}$ is sufficiently large that the
bare oscillator states do not play an important role in determining
the condensation properties for the system.

\subsubsection{Intermediate energy region}

In sufficiently deep lattices many excited bands may be bound by the
lattice, and will contribute to the density of states. In this case
for energies greater than $e_{2}$ and less than $E_{{\rm bare}}$,
the various density of states we have already outlined above will
be inadequate. It is difficult to provide a reliable analytic description
of these excited band contributions for several reasons: (1) Anharmonic
effects of the lattice make predicting the locations (i.e. $e_{nj}$)
of these bands difficult. (2) The tunneling between sites in excited
bands is much larger and worsens the localized state approximation.
This necessitates an effective mass modified harmonic oscillator treatment
(c.f. Eq.  (\ref{eq:eLE})) that crosses over to localized states at higher
energies. Furthermore, large asymmetry between directions can occur
depending on the orientation of the vibrational excitations of each
band, making any form of Debye approximation of limited use.

Here we do not treat these higher bands analytically. For typical
experimental parameters the energy scale of these modes (i.e. $e_{2}$)
is well above $kT_{c}$, and a complete description is not required%
\footnote{From comparisons with full numerical results we have determined spline
interpolating the localized density of states at the top of the first
excited band up to the bare density of states at energy $E_{{\rm bare}}$
provides quite a good description. However, we do not use this here.%
}.

\subsection{Full numerical solution\label{sub:FullNumSoln}}

To test the predictions of this paper we have made a full numerical
solution for the single particle eigenstates of Eq. (\ref{eq:Hsp}).
To do this we use the separability of the Hamiltonian to convert this
eigenvalue problem to a set of three 1D problems. Because the harmonic
potential is quite weak (typically $\omega_{j}\sim0.01-0.05\omega_{R}$
in experiments), a large number of lattice sites need to be represented
to find eigenstates up to a convenient maximum energy (usually $E_{\max}\approx\epsilon_{g}+25E_{R}$),
chosen so that the density of states we construct will be useful for
temperatures up to about $T\sim5E_{R}/k$. We use a planewave decomposition
to represent the eigenstates of the combined potential, chosen because
it provides an efficient representation of the rapidly varying lattice
potential. Typically of order $10^{4}-10^{5}$ planewave modes are
used to represent the several thousand eigenstates in the energy range
of interest. 

For the purposes of comparison to our analytic results, it is useful
to construct a smoothed density of states, defined as \begin{equation}
\bar{g}(\epsilon)=\frac{1}{2\Delta\epsilon}\int_{\epsilon-\Delta\epsilon}^{\epsilon+\Delta\epsilon}d\epsilon\,\sum_{ijk}\delta(\epsilon-[\epsilon_{i}^{(1)}+\epsilon_{j}^{(2)}+\epsilon_{k}^{(3)}]),\label{eq:smoothedDOS}\end{equation}
that gives an average number of eigenstates per unit energy with energies
lying within $\Delta\epsilon$ of $\epsilon$, where $\{\epsilon_{i}^{(j)}\}$
are the (1D) single particle energies ($i=0,1,\ldots$) in the $x_{j}$-
direction obtained form the numerical diagonalization.

\section{Results\label{sec:Results}}

\subsection{Density of states} 
\begin{figure}
\includegraphics[width=3.4in]{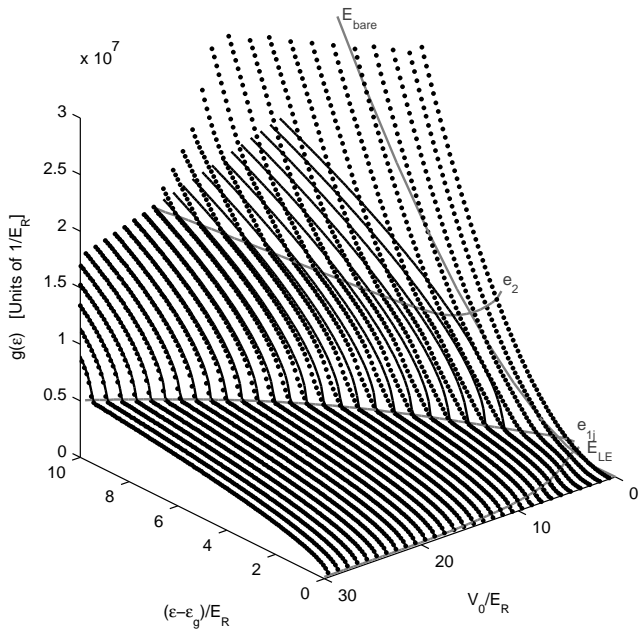}
\caption{\label{fig:numDOScomp}Comparison of numerical smoothed density of
states $\bar{g}(\epsilon)$ (dots) to analytic density of states $\tilde{g}(\epsilon)$
(black lines) for a 3D combined harmonic lattice potential. The lattice
 depth parameters are  the same for each direction,
i.e. $V_{j}=V_{0}$ for $j=1,2,3$. For reference, the characteristic
energy scales $E_{{\rm LE}}$ , $e_{1j}$, $e_{2}$ and $E_{{\rm bare}}$
are shown as gray curves. Isotropic harmonic trap taken with $\bar{\omega}=0.01\omega_{R}$.
All energies are measured relative to $\epsilon_{g}$.}
\end{figure}

Here we investigate the accuracy and applicability of our combined density
of states (\ref{eq:gLE}) and (\ref{eq:gLoc}) by comparison with
the smoothed density of states obtained from the full numerical solution
(see Fig. \ref{fig:numDOScomp}). For definiteness, the analytic density
of states we use is constructed piecewise from results (\ref{eq:gLE})
and (\ref{eq:gLoc}), as \begin{eqnarray}
\tilde{g}(\epsilon) & = & \left\{ \begin{array}{cc}
g_{{\rm LE}}(\epsilon)\quad & e_{0}<\epsilon<E_{{\rm LE}},\\
g_{{\rm Loc}}(\epsilon)\quad & E_{{\rm LE}}\le\epsilon.\end{array}\right.\label{eq:piecewise_gE}\end{eqnarray}
Of course this result can only be expected to furnish a good description
for $\epsilon\lesssim e_{2}.$ In the context of current experiments
this energy range should be sufficiently large that this piecewise
density of states will be useful over a broad parameter regime. E.g.
for $^{87}$Rb in a $10E_{R}$ deep lattice ($a=425$nm), we have
that $(e_{2}-e_{0})\sim10E_{R}\sim k\times1.5\mu$K. 

We make a few observations regarding the results in Fig. \ref{fig:numDOScomp}:
\begin{enumerate}
\item For shallow lattices $E_{{\rm bare}}$ may be sufficiently small that
the transition to bare oscillator states occurs before $E_{{\rm LE}}$
is reached. Indeed, the density of states is mostly harmonic oscillator
like (i.e. $\bar{g}\sim\epsilon^{2}$) for $ $ lattice depths less
than $4E_{R}$, and for this reason the analytic result is only shown
for depths greater than this. 
\item For lattice depths less than $10E_{R}$ the onset of the localized
excited band states in $\tilde{g}(\epsilon)$ for $\epsilon\sim e_{1j}$
is too rapid compared to the numerical results, arising because the
lowest energy states in the excited band are harmonic oscillator like.
However, agreement is observed to improve with increasing lattice
depth, such that for $V_{0}\gtrsim20E_{R}$ the numerical and analytic
results are almost indistinguishable.
\item The energy scale $e_{2}$ increases quite rapidly with lattice depth,
justifying our neglect of additional excited bands in the analytic
density of states.$ $
\end{enumerate}

\subsection{Analytic prediction for the critical temperature}

As is apparent \textbf{}from Fig. \ref{fig:numDOScomp}, for lattices
with $V_{j}\gtrsim4E_{R}$, the majority of the low energy spectrum
is well described by the first term of the localized density of states
(\ref{eq:gLoc}), and thus we use this term to estimate the critical
temperature. 

The total number of particles in the ground band localized states,
as a function of inverse temperature ($\beta=1/kT$ ) and chemical
potential ($\mu$), is given by \begin{eqnarray}
N_{{\rm Loc}}(\beta,\mu) & = & \int_{w_{0}}^{\infty}d\epsilon\,\frac{g_{0}(\epsilon-w_{0})}{e^{\beta(\epsilon-\mu)}-1},\label{eq:Nloc}\\
 & = & \frac{16}{\pi^{2}}\left(\frac{kT\,\omega_{R}}{\hbar\bar{\omega}^{2}}\right)^{\frac{3}{2}}\!\Gamma\left(\!\frac{3}{2}\!\right)g_{\frac{3}{2}}\left(e^{\beta(\mu-w_{0})}\right),\label{eq:Nloc2}\end{eqnarray}
where $g_{s}(z)=\sum_{k=1}^{\infty}z^{k}/k^{s}$ is the polylogarithm
function. 

Following the usual procedure\cite{HuangStatMech} we identify the
critical temperature for condensation by taking the gas to be saturated
($\mu\to w_{0}$) and setting $N_{{\rm Loc}}(\beta_{c0},w_{0})=N$
(the total number of atoms), giving\begin{equation}
T_{c0}\approx\frac{0.4141}{k}\left(\frac{\hbar\bar{\omega}^{2}}{\omega_{R}}\right)N^{2/3},\label{eq:Tc0}\end{equation}
where $\beta_{0c}=1/kT_{c0}$ and $ $we have used that $g_{3/2}(1)=\zeta(3/2)\approx2.612$, with $\zeta(s)=\sum_{k=1}^{\infty}1/k^s$ the Reimann zeta function.
This expression has the same $N^{2/3}$ dependence as the critical
temperature for the uniform Bose gas.

\subsection{Corrections to analytic critical temperature\label{SEC:corrections}}

Expression (\ref{eq:Tc0}) for $ $$T_{c0}$ is based solely on the
localized ground band   states. The effect of the low energy
states (\ref{eq:gLE}) and excited band states (\ref{eq:gLoc}) are
in general significant. We now consider the effect of these on $T_{c0}$
under the assumptions that $(E_{{\rm LE}}-e_{0})/kT_{c0}\ll1$ and
$(e_{2}-e_{0})/kT_{c0}\gg1$. 

\begin{widetext}

\subsubsection{Low energy correction}

The first correction we consider is to account for the low energy
spectrum, described in Sec.\ref{sub:LowE}. To do this we replace $g_{0}(\epsilon-w_{0})$ in (\ref{eq:Nloc})
for $\epsilon<E_{{\rm LE}}$ by the low energy density of states (\ref{eq:gLE}).
This changes $N_{{\rm Loc}}(\beta_{c0},\mu)$ by an amount\begin{eqnarray}
\Delta N_{{\rm LE}} & = & \int_{\epsilon_{0}}^{E_{{\rm LE}}}d\epsilon\,\frac{g_{{\rm LE}}(\epsilon)}{e^{\beta_{c0}(\epsilon-e_{0})}-1}-\int_{w_{0}}^{E_{{\rm L}E}}d\epsilon\,\frac{g_{0}(\epsilon-w_{0})}{\epsilon^{\beta_{c0}(\epsilon-w_{0})}-1},\label{eq:DNLE}\\
 & \approx & kT_{c0}\left[\frac{(E_{{\rm LE}}-e_{0})^{2}}{4\left(\hbar\overline{\omega^{*}}\right)^{3}}-\frac{32}{\pi^{2}}\left(\frac{\omega_{R}}{\hbar\bar{\omega}^{2}}\right)^{3/2}\sqrt{E_{{\rm LE}}-w_{0}}\right]\label{eq:DNLE2}\end{eqnarray}
 where we have assumed that $(E_{{\rm LE}}-e_{0})\ll kT_{c0}$ to
arrive at the last line %
\footnote{Since $E_{{\rm LE}}-e_{0}>E_{{\rm LE}}-w_{0}$ , it also holds that
$(E_{{\rm LE}}-w_{0})\ll kT_{c0}$.%
}.

\subsubsection{Chemical potential correction}

Associated with the change in the low energy density of states is
the change in ground state energy from $w_{0}$ (for the localized
spectrum) to $\epsilon_{g}$ (for the low energy spectrum (\ref{eq:Eg})).
Replacing the saturated chemical potential by the ground state energy,
i.e. setting $\mu\to\epsilon_{g}$ in (\ref{eq:Nloc2}) we obtain
\begin{equation}
\Delta N_{{\rm \mu}}=\frac{-32}{\pi^{3/2}}\left(\frac{\omega_{R}}{\hbar\bar{\omega}^{2}}\right)^{3/2}\sqrt{w_{0}\!-\!\epsilon_{g}}\left(1\!+\!\frac{\zeta(\frac{1}{2})}{2}\sqrt{\frac{w_{0}-\epsilon_{g}}{\pi kT_{c0}}}\right)kT_{c0}.\label{eq:DNFS}\end{equation}
In deriving this result we have assumed that $(w_{0}-\epsilon_{g})/kT_{c0}\ll1,$
so that we can approximate the argument of the polylogarithm as $1-(w_{0}-\epsilon_{g})/kT_{0c}$,
and use the expansion $g_{3/2}(1-x)\approx\zeta(3/2)-2\sqrt{\pi x}-\zeta(1/2)x.$
We note that $\zeta(\frac{1}{2})\approx-1.460$ and the square root
term accounts for the infinite slope of $g_{3/2}(z)$ at $z=1$.

\end{widetext}

\subsubsection{Excited band correction }

As discussed in the derivation of Eq. (\ref{eq:gLoc}), at an energy
scale of $e_{1j}$ excited band states become accessible to the system,
and contribute additional states described by the density of states
$g_{0}(\epsilon-e_{1j})$ . The additional atoms accommodated in these
states at $T_{c0}$ is given by\begin{eqnarray}
\Delta N_{{\rm EB}} & = & \sum_{j=1}^{3}\int_{w_{1j}}^{\infty}d\epsilon\,\frac{g_{0}(\epsilon-e_{1j})}{\epsilon^{\beta_{c0}(\epsilon-w_{0})}-1},\label{eq:DNEB}\\
 & \approx & \sum_{j=1}^{3}\frac{8}{\pi^{3/2}}\left(\frac{kT_{c0}\omega_{R}}{\hbar\bar{\omega}^{2}}\right)^{3/2}e^{-(e_{1j}-w_{0})/kT_{c0}},\nonumber \end{eqnarray}
where we have taken $(e_{1j}-w_{0})\gg kT_{c0}$. Note that in calculating
this term we have summed over all contributing first excited bands.

\subsubsection{Corrected critical temperature}

Combining all the above results we arrive at a new estimate for the
transition temperature. To do this we set\begin{equation}
N=N_{{\rm Loc}}(\beta_{c1},w_{0})+\Delta N_{{\rm LE}}+\Delta N_{\mu}+\Delta N_{{\rm EB}},\label{eq:Ncorrect}\end{equation}
where $\beta_{c1}=1/kT_{c1}$ is the corrected transition temperature.
Assuming that $|T_{c1}-T_{c0}|\ll T_{c0}$, we obtain\begin{equation}
T_{c1}\approx T_{c0}\left[1-\frac{2}{3}(\Delta N_{{\rm LE}}+\Delta N_{{\rm \mu}}+\Delta N_{{\rm EB}})/N\right],\label{eq:Tc1}\end{equation}
to first order in the $\Delta N$-corrections. The validity conditions
are, as stated above, that $(E_{{\rm LE}}-e_{0})/kT_{c0}\ll1$ and
$(e_{2}-e_{0})/kT_{c0}\gg1$. This will ensure that all the changes
($\Delta N$) are small compared to $N$, however we caution that
sometimes due to cancellation a particular $\Delta N$ can be small
even when the validity condition is not satisfied.

We make the following observations on these corrections:
 \begin{itemize}
\item[$\Delta N_{{\rm LE}}$:] The low energy density of states tends to
increase much more slowing from its zero point than the localized
density of states does. Thus in replacing $g_{0}(\epsilon)$ by $g_{{\rm LE}}(\epsilon)$
in (\ref{eq:Nloc}), the number of states at low energy and hence
the number of atoms in the saturated thermal cloud both decrease.
This leads to an increase in the critical temperature.
\item[$\Delta N_{{\rm \mu}}$:] The downward shift of the chemical potential
when we change the saturated chemical potential from $w_{0}$ to $\epsilon_{g}$
leads to a decrease in the number of atoms in the saturated thermal
cloud, and hence an increase in the critical temperature.
\item[$\Delta N_{{\rm EB}}$:] Including higher bands brings additional
states and hence increases the number of atoms in the saturated thermal
cloud. This has the effect of decreasing the critical temperature.
\end{itemize}

Interestingly the dominant corrections at low temperatures ($\Delta N_{{\rm LE}}$
and $\Delta N_{{\rm \mu}}$) both lead to an increase in $T_{c}$,
whereas the dominant correction at higher temperatures ($\Delta N_{{\rm EB}}$)
shifts $T_{c}$ downwards.

\subsection{Numerical calculations of $T_{c}$ }

While the analytic calculation provides a useful critical temperature
estimate, the complexity of the spectrum in the combined harmonic-lattice
potential necessitates a numerical solution. Here we discuss our procedure
for calculating the critical temperature using the spectrum determined
by full numerical diagonalization of (\ref{eq:Hsp}) and give a simple
numerical scheme that makes use of the piecewise density of states we have
developed in Secs. \ref{sub:LowE} and \ref{sub:Loc}.

\subsubsection{Full numerical calculation\label{sub:FullNumCalc}}

From the results of our full diagonalization we determine the one-dimensional
energy spectrum $\{\epsilon_{i}^{(j)}\}$ over a large energy range, typically
including all states up to energy $25E_{R}$ above the 1D ground state
energy (as discussed in Sec. \ref{sub:FullNumSoln}). The thermal
properties of the system are then calculated over a temperature range
by   iterating the chemical potential $\mu$ to find the
desired total number of atoms, i.e. root-finding the expression$f(\mu)=\left[\sum_{ijk}\{\exp([\epsilon_{i}^{(1)}+\epsilon_{j}^{(2)}+\epsilon_{k}^{(3)}-\mu]/kT)-1\}^{-1}-N\right]$
for each $T.$ From this calculation we hence evaluate the condensate
population as a function of temperature, i.e, $N_{0}(T)=\{\exp([\sum_{j=1}^{3}\epsilon_{0}^{(j)}-\mu]/kT)-1\}^{-1}$,
and determine the condensation temperature as that at which $|(\partial N_{0}/\partial T)/N_{0}|$
(i.e. the relative change in the ground state occupation) is maximised.

\subsubsection{Simple numerical calculation\label{sub:TcN}}

The critical temperature can also be estimated by performing a simple
numerical integral using the piecewise analytic density of states
(\ref{eq:piecewise_gE}) under the saturated thermal cloud condition
($i.e.$ $\mu\to e_{0}$)

\begin{equation}
N(T)=\int_{e_{0}}^{E_{\max}}d\epsilon\frac{\tilde{g}(\epsilon)}{e^{(\epsilon-e_{0})/kT}-1}.\label{eq:NTN}\end{equation}
This result can then be numerically inverted to give a critical temperature
estimate $T_{cN}=T_{cN}(N).$ The energy $E_{\max}$ appearing in
the integral has to be chosen such that $E_{\max}\gg kT$, in which
case the result will be independent of $E_{\max}$. 

This approach is significantly simpler than the full numerical calculation
because it does not require a full numerical diagonalization. Indeed
the information needed for $\tilde{g}(\epsilon)$ can be obtained
from results of the uniform lattice or tight-binding approximations,
as discussed in Sec. \ref{sub:EnergyScales}.

\subsection{Comparison of analytic and numerical critical temperatures}

\begin{figure}
\includegraphics[width=3.4in]{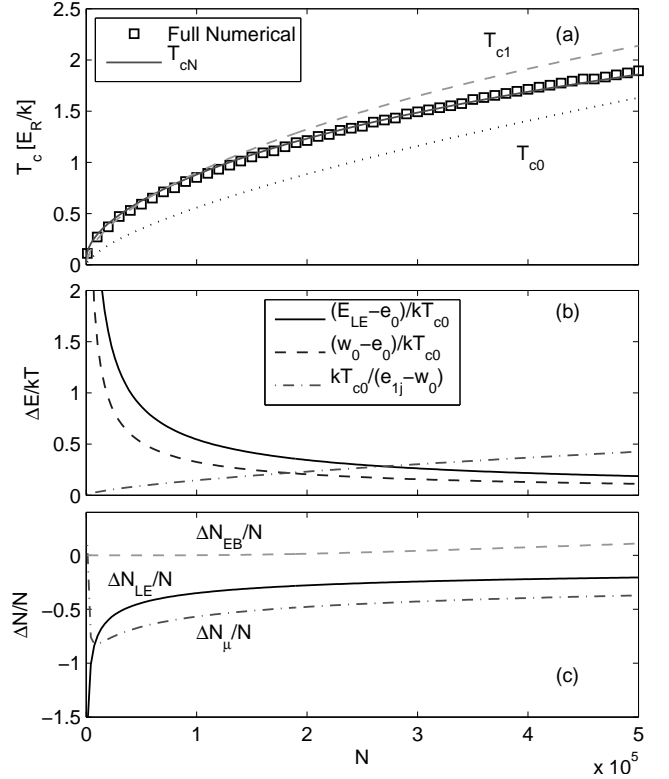}
\caption{\label{Fig:Tc} Comparison of analytic and numerical critical temperatures
(a) Full numerical results for $T_{c}$ (squares), analytic results $T_{c0}$ (dotted), $T_{c1}$ (dashed) and simple numerical result using piecewise analytic density of states $T_{cN}$ (solid).
(b) Energy scales compared to $T_{c0}$. (c) $\Delta N$-corrections.
Calculation parameters: isotropic harmonic trap with  $\bar\omega=0.025\omega_{R}$, and lattice  depth parameters $V_{j}=8E_R$ for $j=1,2,3$.}
\end{figure}

In Fig. \ref{Fig:Tc}(a) we show  analytic and numerical results  for the critical temperature. To relate these parameters
to those in experiment we note that for $^{87}$Rb in a lattice with
$a=425$nm, the trap frequency corresponds to $\omega\approx2\pi\times31\, s^{-1}$
while the temperature scale is $E_{R}/k\approx152nK$. These results show the general behaviour
we have observed over a wide parameter regime. $T_{c0}$ provides
a useful critical temperature estimate, though is noticeably shifted
relative to the full numerical result. Including first order corrections
$T_{c1}$ provides a quantitatively much more accurately result, although
its agreement with the full numerical result worsens for large $N$.
Interestingly the simple numerical result $T_{cN}$ outlined in Sec.
\ref{sub:TcN} provides an accurate description over the full range
considered.

In Fig. \ref{Fig:Tc}(b) and (c) we explore the validity conditions
for our derivation of the critical temperature. Note for the potential
parameters used for the results in Fig. \ref{Fig:Tc} we have that
$E_{{\rm LE}}-e_{0}=0.304E_{R}$, $E_{{\rm LE}}-w_{0}=0.123E_{R}$,
$e_{1j}-w_{0}=3.83E_{R}$, and $e_{2}-e_{0}=6.7E_{R}$. The relative
size of the parameters $(E_{{\rm LE}}-e_{0})/kT_{c0}$, $(E_{{\rm LE}}-w_{0})/kT_{c0}$
and $kT_{c0}/(e_{1j}-w_{0})$ are shown in Fig. \ref{Fig:Tc}(b).
We require all of these parameters to be small for our analytic calculation
to be valid. These results show that for small $N$ the critical temperature
is sufficiently low that a first order treatment of the low energy
spectrum is not appropriate (i.e. both $(E_{{\rm LE}}-e_{0})/kT_{c0}$
and $(E_{{\rm LE}}-w_{0})/kT_{c0}$ are large). 

At larger atom numbers ($N$) the term $kT_{c0}/(e_{1j}-w_{0})$ tends
to grow reflecting the increased importance of excited band states.
In Fig. \ref{Fig:Tc}(c) we show the related values of $\Delta N_{{\rm LE}}$,
$\Delta N_{\mu}$ and $\Delta N_{{\rm EB}}$. At small $N$ ( and
hence small $T_{c0}$) the expansions we have used to obtain$\Delta N_{{\rm LE}}$
and $\Delta N_{\mu}$ are not valid. As $N$ increases these contributions
become less significant relative to $N,$ however as $\Delta N_{\mu}$
scales like $\sqrt{(E_{{\rm LE}}-w_{0})/kT_{c0}}$ it decreases rather
slowly with increasing $T_{c0}$. Finally, the excited band contribution
becomes gradually more significant with increasing number.

\section{General behaviour of condensation in the combined potential\label{sec:GenRes}}

\begin{figure}
\includegraphics[width=3.5in,keepaspectratio]{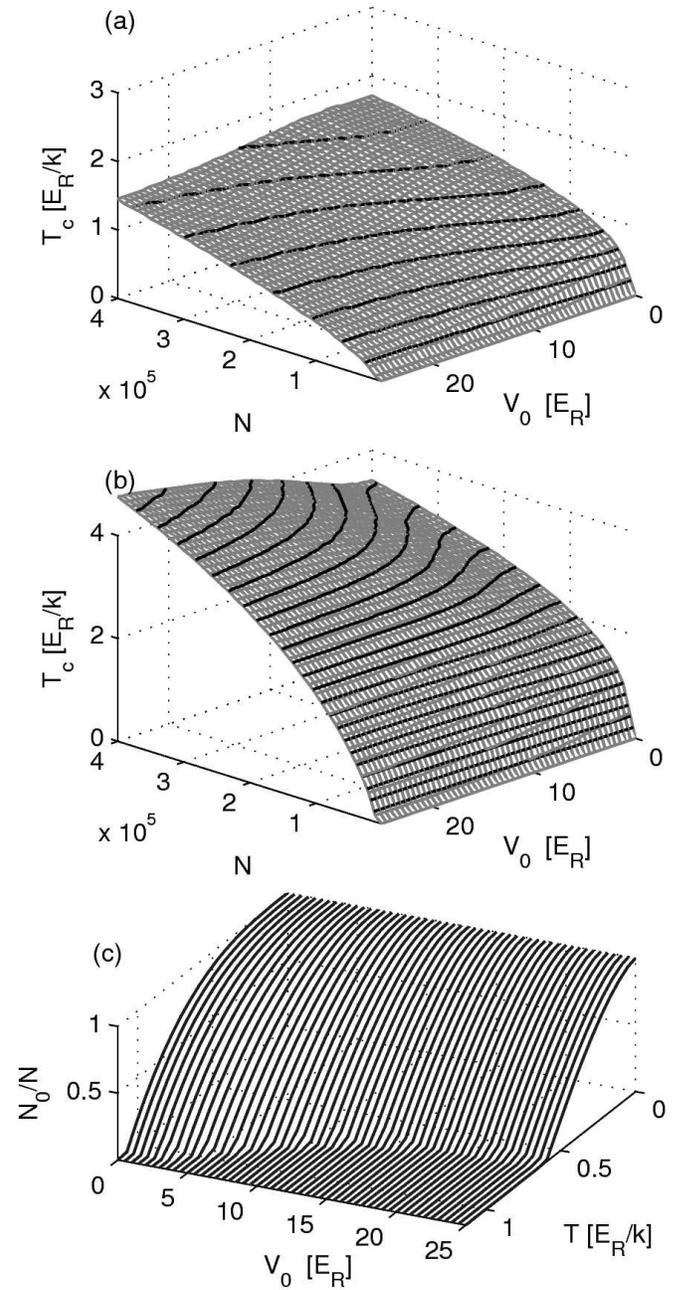}
\caption{\label{fig:CondFig} Bose-Einstein condensation in a combined harmonic
lattice potential. (a) Critical temperature as a function of lattice
depth (all $V_{j}=V_{0}$) and total atom number for an isotropic
harmonic trap with $\bar{\omega}=0.025\omega_{R}$. Isothermal levels
spaced by $0.2E_{R}/k$ shown as contour lines. (b) As for (a) but
with $\bar{\omega}=0.05\omega_{R}$. (c) Condensate fraction versus
temperature for the same parameters as (a) and $N=1\times10^{5}$
atoms.}
\end{figure}

In figure \ref{fig:CondFig} we show the results of our full numerical
calculation (as discussed in Sec. \ref{sub:FullNumCalc}) for the
critical temperature and condensate fraction over a wide parameter
regime. For the case of $\bar{\omega}=0.025\omega_{R}$ in Fig. \ref{fig:CondFig}(a)
we see that as the lattice depth increases the critical temperature
of the system decreases. While for the case of $\bar{\omega}=0.05\omega_{R}$
shown in Fig. \ref{fig:CondFig}(b) the critical temperature instead
tends to increase with increasing lattice depth (for $N$ sufficiently
large). 

To understand these results we recall the critical temperature for
a harmonically trapped gas \begin{equation}
T_{{\rm harm}}=\frac{\hbar\bar{\omega}}{k}\left(\frac{2N}{\Gamma(3)\zeta(3)}\right)^{\frac{1}{3}}.\label{eq:Tharm}\end{equation}
In comparison to our analytic result given in Eq. (\ref{eq:Tc0}),
we note that the critical temperature for the combined potential scales
with mean trap frequency and total atom number at higher powers, i.e.
$\bar{\omega}^{2}$ and $N^{2/3}$ respectively. Locating the trap
frequency at which the critical temperatures for the harmonic and
combined potentials are the same determines a critical mean trap frequency
($\bar{\omega}_{c}(N)$):\begin{equation}
\frac{\bar{\omega}_{c}(N)}{\omega_{R}}=\frac{4}{\pi}\left(\frac{\zeta\left(\frac{3}{2}\right)^{2}}{\zeta(3)}\right)^{\frac{1}{3}}\frac{1}{\sqrt[3]{N}}.\label{eq:trapc}\end{equation}
For $\bar{\omega}>\bar{\omega}_{c}$ the critical temperature is higher
in the combined potential than for the pure harmonic trap, whereas
for $\bar{\omega}<\bar{\omega}_{c}$ the pure harmonic potential has
a higher critical temperature. For $N=10^{5}$ atoms we find that
$\bar{\omega}_{c}\sim0.049\omega_{R}$, which is consistent with Figs.
\ref{fig:CondFig}(a) and \ref{fig:CondFig}(b) which lie either side
of this value. Since $\bar{\omega}_{c}$ is based on the simple critical
temperature estimate (\ref{eq:Tc0}), it will only be valid for cases
where the critical temperature is not too high or low (as given by
the validity conditions in Sec. \ref{SEC:corrections}).

In Fig. \ref{fig:CondFig}(c) we show the condensate fraction versus
temperature for a system of $10^{5}$ atoms in a combined potential with $\bar{\omega}=0.025\omega_{R}$. As the lattice depth increases the critical temperature
shifts downwards (as can also be discerned from Fig. \ref{fig:CondFig}(a)),
and the characteristic shape of the condensate fraction dependence
on temperature, $ $$N_{0}\sim[1-(T/T_{c})^{\alpha}]$, changes from
$\alpha\sim3$ to $\alpha\sim3/2$. These predicted features should be verifiable
by current experiments.

\section{Conclusion}
We have performed a comprehensive study of the critical temperature
for an ideal Bose gas in a combined harmonic lattice potential. We
have described distinctive regions of the spectrum and have shown
that a simple piecewise density of states provides an accurate characterization
of this system for lattice depths greater than about $4E_R$. We have developed an analytic expression for the critical
temperature in the combined potential. The corrections to this result
are typically significant, and we have shown that including them provides
a useful estimate for the critical temperature obtained by a full
numerical calculation. Additionally, we give a simple numerical procedure
based on piecewise density of states that provides an accurate prediction
for the critical temperature. Finally we have presented results over
a wide parameter regime appropriate to current experiments and have
shown that the critical temperature in the combined potential can
be increased or decreased relative to that of the pure harmonic trap.

\section*{Acknowledgments}
PBB would like to thank the University of Otago and the Marsden Fund
of New Zealand for financial support. WXW would like to acknowledge financial support from the China Scholarship Council under grant 2004837076.
Valuable discussions with Patrick Ledingham and Emese Toth are gratefully acknowledged.

\bibliographystyle{apsrev}
\bibliography{BoseLatt_Bib}

\end{document}